\documentclass{article}
\usepackage{spconf,amsmath,graphicx}
\usepackage{setspace}
\usepackage{amssymb}
\usepackage{float}
\usepackage{booktabs}
\usepackage{multirow}
\title{An Efficient Temporary Deepfake Location Approach Based Embeddings for Partially Spoofed Audio Detection}
\name{Yuankun Xie, Haonan Cheng, Yutian Wang, Long Ye\sthanks{Long Ye is corresponding author.}}
\address{
	State Key Laboratory of Media Convergence and Communication, \\
	Communication University of China, Beijing 100024, China}

\begin{document}
\ninept
\maketitle
\begin{abstract}
Partially spoofed audio detection is a challenging task, lying in the need to accurately locate the authenticity of audio at the frame level. To address this issue, we propose a fine-grained partially spoofed audio detection method, namely Temporal Deepfake Location (TDL), which can effectively capture information of both features and locations. Specifically, our approach involves two novel parts: embedding similarity module and temporal convolution operation. To enhance the identification between the real and fake features, the embedding similarity module is designed to generate an embedding space that can separate the real frames from fake frames. To effectively concentrate on the position information, temporal convolution operation is proposed to calculate the frame-specific similarities among neighboring frames, and dynamically select informative neighbors to convolution. Extensive experiments show that our method outperform baseline models in ASVspoof2019 Partial Spoof dataset and demonstrate superior performance even in the cross-dataset scenario.
\end{abstract}
\begin{keywords}
partially spoofed audio detection, temporal deepfake location, embedding learning.
\end{keywords}
\section{Introduction}
AI generated content (AIGC) technology has witnessed swift progress in recent years, particularly in speech-related applications like text-to-speech (TTS) \cite{huang2022prodiff,tan2022naturalspeech,wang2023neural} and voice conversion (VC) \cite{chan2022speechsplit2,chen2021again,tang2022avqvc}. Although these technologies have brought about convenience, they have also posed significant security threats. Thus, various initiatives and challenges, such as ASVspoof \cite{nautsch2021asvspoof, delgado2021asvspoof}, have been established to foster research on countermeasure solutions that safeguard speech applications and human listeners against spoofing attacks. Nevertheless, a significant scenario has been overlooked in most datasets and challenges where a bonafide speech utterance is contaminated by synthesized speech segments, leading to partial spoofing (PS). Attackers can use PS to alter sentence semantics, and such modifications can be easily accomplished at low cost. For instance, attackers can easily modify single word such as time, place, and characters in sentence to dramatically change the semantics. Furthermore, If attackers have knowledge of phonology, they can manipulate vowels and even consonants such as “pan,”“pin,”“pen,” which are smaller than the word level. Therefore, defending against such fine-grained PS scenarios poses significant challenges for defenders.

In recent years, there are several studies about PS scenarios for Audio Deepfake Detection (ADD).  Yi et al. \cite{yi21_interspeech} create a dataset that focuses on changing a few words in an utterance for half-truth audio detection. At the same time, Zhang et al. \cite{zhang21ca_interspeech} construct a speech database called `PartialSpoof' designed for PS scenarios. The above two datasets are the beginning of the research for PS scenario in ADD task. Afterward, Zhang et al. \cite{zhang2021multi} propose the SELCNN network to enhance the ability of the accuracy of the utterance. Lv et al. \cite{lv2022fake} use Wav2Vec2 (W2V2) \cite{baevski2020wav2vec} as front-end, ECAPA-TDNN \cite{desplanques2020ecapa} as back-end achieving the first rank in ADD 2022 Track 2\cite{yi2022add}. Although the above research shows effectiveness at the utterance level detection in PS, they do not pinpoint specific segments with precision. Thus, Zhang et al. \cite{zhang2022partialspoof} extended the previous utterance-level PS dataset labels to frame-level and proposed corresponding W2V2-based countermeasures to enhance frame-level detection capability.
\begin{figure*}[!t]
	\centering
	\includegraphics[width= 5in]{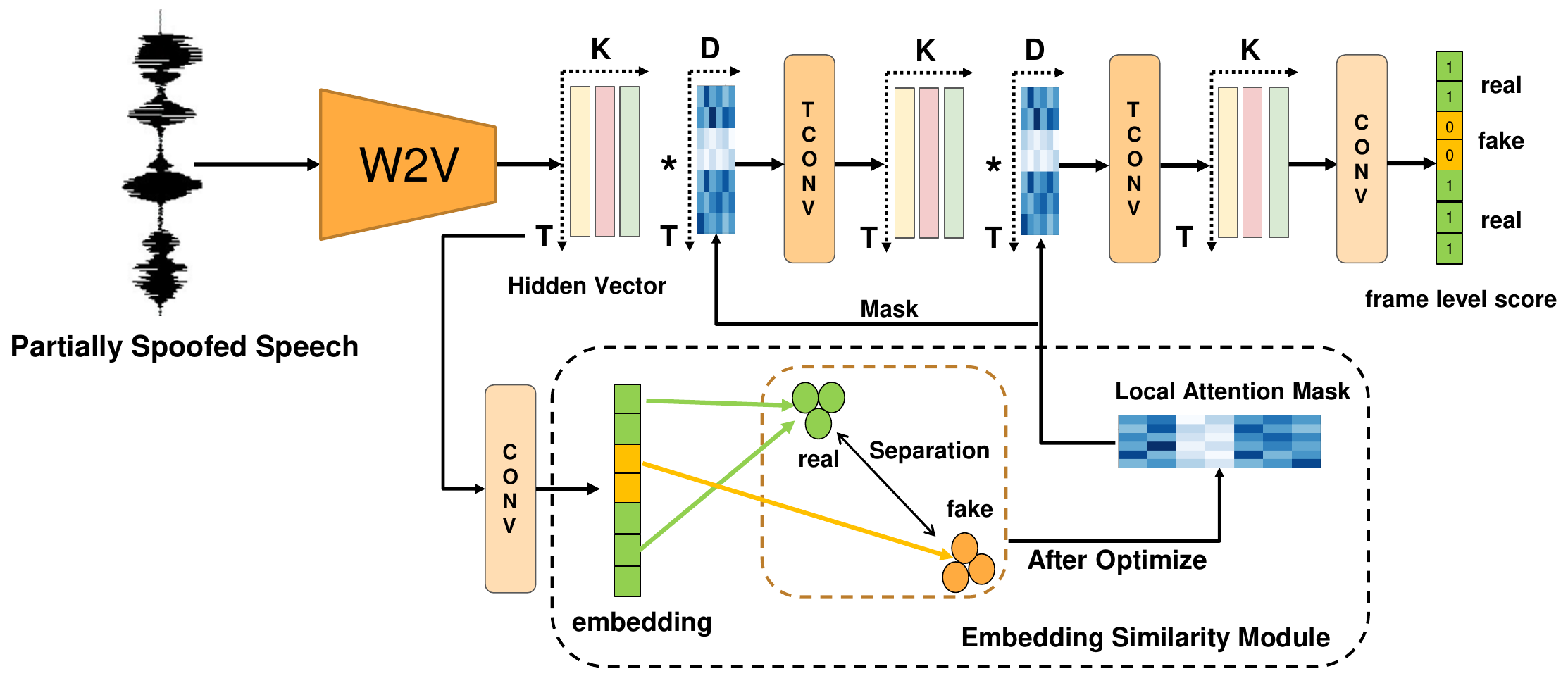}
	\hfil
	\caption{The entire structure of our proposed Temporal Deepfake Location (TDL) method.}
	\label{fig:pipeline}
\end{figure*}

The aforementioned methods solely utilize existing ADD models such as LCNN, currently lacking specific approaches tailored to the PS scenario, particularly in terms of precise frame-level localization. To address this challenge, we propose a novel Temporal Deepfake Location (TDL) method. For front-end, we take advantage of W2V2 \cite{babu2021xls}.  By training on a vast corpus of genuine speech from diverse source domains, W2V2 can effectively discriminate the real and fake in complex acoustic scenarios. For back-end, our primary focus is on fine-grained locating the genuine and spoofed speech segment. To clearly distinguish the real and fake in feature level, we first design the embedding similarity module to separate the real and fake frames in embedding space and get a high-quality embedding similarity vector. Then, we propose temporal convolution operation to locate the region from the embedding vector. The local similarity for each temporal position is calculated from the embedding. By this means, we can obtain a frame-specific weight to guide the convolution making a temporal sensitive calculation. Our main contributions can be summarized as follows:
\begin{itemize}
	\item We propose TDL method, an efficient and effective ADD method for PS scenarios which combines a embedding similarity module and temporal convolution operation to effectively capture both feature and positional information.
	\item The proposed method outperforms baseline models in ASV spoof 2019PS dataset and demonstrate superior performance even in cross-dataset experiments.
\end{itemize}

\section{Proposed Method}
\subsection{Problem statement and overview} 
In PS scenarios, the fake audio segment is inserted within the genuine speech. Our target is to detect the real and fake segments at frame level. Given the large-scale self-supervised audio feature 
$f = (f_{1},f_{2},...f_{T})\in R^{D \times T}$, where $D$ and $T$ denote the dimension of audio feature and the number of frames respectively. The whole task is defined as input feature $f$ and output the frame level label $ y = (y_{1},y_{2},...y_{T})\in \{0,1\}^{T}$, where 1 represents the real frames and 0 represents the fake frames.

The framework of our proposed TDL is depicted in Figure \ref{fig:pipeline}. First, we utilize Wav2Vec-XLS-R to extract the frame level feature from the raw audio. Then, for enhanced identification of genuine and fake distinctions at the embedding level, we devise an embedding similarity module to segregate authentic and synthetic frames within the embedding space. Next, to capture the position information, we adopt temporal convolution operation by focusing on frame-specific similarities among neighboring frames. Finally, we employ 1D convolutional layers and fully connected layers for downsampling to the frame level label to compute the Binary Cross-Entropy (BCE).

\subsection{W2V2 front-end}
W2V2 based front-end is trained by solving a contrastive task over a masked feature encoder. Firstly, speech signals in various lengths are passed through a feature extractor consisting of seven convolutional neural network (CNN) layers. Subsequently, context representations are obtained using a Transformer network \cite{vaswani2017attention} comprising of 24 layers, 16 attention heads, and an embedding size of 1024. In practice, we utilize the Hugging Face version of wav2vec2-XLS-R-300M\footnote{https://huggingface.co/facebook/wav2vec2-xls-r-300m} and freeze the weights of the front-end. The front-end model is pre-trained with 436k hours of unannotated genuine speech data in 128 languages. Consequently, the last hidden states from the transformer can effectively represent the contextualized information of genuine speech which is different from the partially fake speech.

\subsection{Embedding similarity module}
To better capture feature-level information, we first distinguish the real and fake frames in the embedding space. Specifically, the W2V2 features are fed into a CONV module, consisting of two sequential 1D-CNNs, which downsamples the embedding dimension from 1024 to 32. The embedding vector is $L2$-normalized. Then we get a embedding vector $ e = (e_{1},e_{2},...e_{T})\in R^{D \times T}$. In the embedding similarity module, we utilize cosine similarity to measure the similarity of two embedding vector $e_{u}$ and $e_{v}$ as follows:
\begin{equation}
	\mathcal{S}\left(\mathbf{e}_{u}, \mathbf{e}_{v}\right)=\frac{\mathbf{e}_{u}^{T} \cdot \mathbf{e}_{v}}{\left\|\mathbf{e}_{u}\right\|_{2} \cdot\left\|\mathbf{e}_{v}\right\|_{2}}.
\end{equation}

To increase the distance between genuine and fake frames in the embedding space and improve generalizability, we computed the cosine similarities between genuine frames, between fake frames, and between genuine and fake frames. Specifically, we ensured that genuine frames from different positions exhibited similarity, fake frames from different positions exhibited similarity, while genuine and fake frames are dissimilar to each other. 

Thus, $\mathcal{L}_{ESM}^{Real}$ and $\mathcal{L}_{ESM}^{Fake}$ are proposed to make the real frames and fake frames in different positions similar:
 \begin{equation}
 	\mathcal{L}_{\mathrm{ESM}}^{\mathrm{Real}}=\max _{\forall e_{x}, e_{y}, x \neq y}\left\lfloor\tau_{\text {same }}-\mathcal{S}\left(\mathbf{e}_{x}, \mathbf{e}_{y}\right)\right\rfloor_{+}, 
 \end{equation}
\begin{equation}
	\mathcal{L}_{\mathrm{ESM}}^{\mathrm{Fake}}=\max _{\forall e_{m}, e_{n}, m \neq n}\left\lfloor\tau_{\text {same }}-\mathcal{S}\left(\mathbf{e}_{m}, \mathbf{e}_{n}\right)\right\rfloor_{+}, 
\end{equation}
where $e_{x}$ and $e_{y}$ refer to distinct positions of real frames, while $e_{m}$ and $e_{n}$ refer to those of fake frames. $\tau_{\text {same }}$ is the similarity threshold between frames from the same category, $\left\lfloor\\ \dots \right\rfloor_{+}$ represents clipping below at zero. It is noteworthy that although we know the positions of frame-level authenticity labels, the temporal dimension of W2V2-XLS-R features does not inherently align with these frame-level labels. To tackle this issue, we ascertain the temporal authenticity in the time dimension of the embedding vector by calculating the ratio between the temporal dimensions of the label and the embedding vector.

$\mathcal{L}_{ESM}^{Diff}$ is proposed to separate the real and fake frames, which can be formulated as:
\begin{equation}
	\mathcal{L}_{\mathrm{ESM}}^{\mathrm{Diff}}=\max _{\forall e_{r}, e_{f}}\left\lfloor\mathcal{S}\left(\mathbf{e}_{r}, \mathbf{e}_{f}\right)-\tau_{\text {Diff }}\right\rfloor_{+}, 
\end{equation}
where $e_{r}$ and $e_{f}$ refer to the embedding vector of real frames and fake frames. $\tau_{\text {diff }}$ is the similarity threshold to constraint the distance between real and fake frames.
Finally, the embedding similarity module is optimized by $\mathcal{L}_{ESM}$, which takes into account the three aforementioned losses in a joint manner. The $\mathcal{L}_{ESM}$ is calculated as follows:
\begin{equation}
	\mathcal{L}_{ESM} = \mathcal{L}_{ESM}^{Real} + \mathcal{L}_{ESM}^{Fake} + \mathcal{L}_{ESM}^{Diff}.
\end{equation}

\subsection{Temporal convolution operation}
To effectively capture the positional information, we use the embedding vector as an local attention mask to perform temporal convolution operations. Consider a audio feature ${\mathbf{X}}\in R^{D_{in} \times T}$, where $D_{in}$ and $T$ represent the dimension of the vector and number of frames respectively. The temporal convolution layer learns a dynamic convolution kernel $\Bbbk\in R^{k \times D_{in} \times D_{out}}$, where $k$ is the size of temporal kernel, $D_{out}$ is the dimension of output feature. We only utilize the dynamic kernel $\Bbbk^{m}\in R^{k \times D_{in}}$ to compute $m^{th}$ channel of the output for convenient. Thus, the temporal convolution operation for the $t^{th}$ feature can be expressed as:
\begin{equation}
	f_{t}^{m}=\sum_{i=0}^{k-1} \mathcal{\Bbbk}^{m}[i,:] \cdot \overline{\mathbf{X}}\left[:, t-\frac{k}{2} +i\right],
\end{equation}
where $f_{t}^{m}$ is the value in the $m^{th}$ channel of output feature vector, $[\cdots]$ means a slice of a matrix, $(\cdot)$ denotes the inner product. $\overline{\mathbf{X}}$ is the modulated feature processed by neighbor similarity calculation:
\begin{equation}
	\begin{aligned}
		&\overline{\mathbf{X}}\left[:, t-\frac{k}{2}+i\right]=
		\mathbf{X}\left[:, t-\frac{k}{2}+i\right] \times \mathbf{a}[i, t], \\&i \in[0, \ldots, k-1],
	\end{aligned}
\end{equation}
where matrix $\mathbf{a}\in R^{k \times T}$ is a similarity matrix that calculate the local similarity for each temporal position, $\mathbf{a}[i, t]$ indicates the similarity between the $t^{th}$ feature vector and its $k$ neighbors.

In practice, we determine the dynamic kernel weight based on the embedding vector generated by ESM module. We apply temporal convolution operation to the W2V2 features on two sequence 1D-CNNs, where both input channel and output channel remain unchanged to maintain consistency in temporal dimension.
\vspace{-0.3cm}
\subsection{Total loss}
Following two consecutive temporal convolution operation layers, to capture additional temporal information and align with the label dimensions, we subsequently employ 1D-CNN, fully connected (FC) layers, and sigmoid activation functions to calculate the BCE loss. The architecture details of TDL is shown in Table \ref{tab:TDL}. The total loss is defined as follow:
\begin{equation}
	\mathcal{L}_{all} = \mathcal{L}_{BCE} + \lambda \mathcal{L}_{ESM},
\end{equation}
where $\lambda$ is set to 0.1 to balance the value of two losses.
\begin{table}[t]
	\caption{Architecture of TDL network.}
	\centering
	\renewcommand{\arraystretch}{0.5}
	\begin{tabular}{cccc}
		\toprule
		module &kernel/stride &output shape\\
		\midrule
		W2V2 &-&(batch,1024,1050) \\
		\midrule
		\multirow{2}{*}{CONV} &3/1&(batch,512,1050) \\
		& 3/1&(batch,32,1050) \\
		\midrule
		TCONV &3/1&(batch,1024,1050) \\
		\midrule
		TCONV &3/1&(batch,1024,1050) \\
		\midrule
		CONV  &1/1&(batch,2,1050) \\
		\midrule
		Flatten/FC &-&(batch,132) \\
		\bottomrule	
	\end{tabular}
	\label{tab:TDL}
\end{table}

\section{Experiments}

\subsection{Database}
Our experiments for PS scenario include two public datasets: ASVspoof2019PS (19PS) \cite{zhang21ca_interspeech} and LAV-DF \cite{cai2022you}. 19PS is constructed based on the ASVspoof2019 LA database \cite{wang2020asvspoof}. All experiments on the 19PS dataset are conducted using 160ms resolution labels. The training, validation, and testing sets are distributed according to the original dataset allocation, consisting of 25,380, 24,844, and 71,237 utterances respectively.

To evaluate the model's generalizability, we conduct additional testing of the 19PS-trained model using the LAV-DF test set. LAV-DF represents a multi-modal temporal forgery dataset, containing a total of 26,100 videos in test set. We extract the audio track of each video and create 160ms frame level genuine and fake labels. 

We calculated the percentage of samples belonging to fake class at both the frame and sentence levels, as shown in the Table \ref{tab:percentage}. We can observe that the frame-level labels in 19PS are balanced, facilitating model training. However, the LAV-DF dataset exhibits a lower proportion of spoof segments, making it unbalanced and presenting greater challenges for detection.
\begin{table}[t]
	\caption{Percentages(\%) of fake class in each dataset.}
	\centering
	\setlength{\tabcolsep}{6pt}
	\renewcommand{\arraystretch}{0.5}
	\begin{tabular}{cccc}
		\toprule
		dataset &subset&frame-level &utterance-level\\
		\midrule
		19PS &train &53.00&89.83\\
		\midrule
		19PS &dev &52.31&89.74\\
		\midrule
		19PS &test &48.03&89.68\\
		\midrule
		LAV-DF &test&10.01&48.82 \\
		\bottomrule
	\end{tabular}
	\label{tab:percentage}
\end{table}

\subsection{Implementation details}
In order to address the issue of variable-length audio inputs, we employ the technique of zero-padding to the maximum length of training set. For the frame of genuine speech, we set the label to one, while for spoofing frame, the label is set to zero. In the case of 19PS, the maximum duration of speech in the training set is 21.03 seconds with a W2V2 feature dimension of (1050,1024) and the number of frames at a resolution of 160 ms is 132. For LFCC, we extracted 60-dimensional LFCC with a combination of static, delta and delta-delta coefficients.

For training strategy, the Adam optimizer is adopted with $\beta_{1}= 0.9$, $\beta_{2}= 0.999$, $\varepsilon$ = $10^{-9}$ and weight decay is $10^{-4}$. We train all of the models for 100 epochs. The learning rate is initialized as $10^{-5}$ and halved every 5 epochs. It is worth mention that no data augmentation method is used for experiment.
\subsection{Evaluation metrics}
In our experiment, we employ four evaluation metrics to assess model performance: Equal error rate (EER), precision, recall, and $F_{1}$ score. All metrics are computed based on frame-level authenticity labels of the partially spoofed audio. Precision, recall, and $F_{1}$ score are defined as follow:
\begin{equation}
	Precison=\frac{T P}{T P+F P},
\end{equation}
\begin{equation}
	Recall=\frac{T P}{T P+F N},
\end{equation}
\begin{equation}
	{F_{1} score}=\frac{2 \cdot Precison \cdot Recall}{Precison+Recall},
\end{equation}
where $TP$, $TN$, $FP$, $FN$ represent the numbers of true positive, true negative, false positive, and false negative samples, respectively. In practice, we employed point-based binary-classification precision, recall, and $F_{1}$ score from Sklearn. Before any evaluation, zero-padding is eliminated based on the actual length of the features.

\section{Results and discussions}
\subsection{Results}
\noindent \textbf{Results on 19PS.} We compare the performance of several baseline models in terms of EER metric, as presented in Table \ref{tab:eerps}. All models are trained on the 19PS training dataset. TDL (w/o ESM) represents our model without ESM module. As shown in Table \ref{tab:eerps}, our model achieve the lowest EER 7.04\% in partially spoofed audio detection task.

Based on the experimental results, We first observe that the impact of feature is greater than backbone. For instance, as seen in first and third row in Table \ref{tab:eerps}, where the backbone is LCNN-BLSTM, the utilization of W2V2 features resulted in a 6.84\% EER decrease compared to LFCC. Conversely, when feature remain consistent, as demonstrated in first and second row of the in Table \ref{tab:eerps}, both employing the shared LFCC attribute, SELCNN-LSTM exhibited a marginal EER reduction of 0.28\% in comparison to LCNN-LSTM. Furthermore, we find that the architecture design of the TDL network aligns well with partial spoofed detection. Specifically, when the features utilized W2V2-XLS-R, the TDL (without ESM module) still exhibits a 1.08\% reduction in EER compared to the LCNN-BLSTM.

\noindent \textbf{Results on LAV-DF.} To validate the generalizability of our proposed model, we train on 19PS and evaluate on the test set of LAV-DF for 4 evaluation metrics. The results of the testing are presented in the Table \ref{tab:lavdf}. Although LAV-DF is an unbalanced dataset, our proposed model achieve the best performance of 11.23\% EER compared to baseline models. 
\begin{table}[t]
	\caption{EER results (\%) on ASVspoof2019 PS dataset.}
	\centering
	\setlength{\tabcolsep}{10pt}
	\renewcommand{\arraystretch}{0.5}
	\begin{tabular}{ccc}
		\toprule
		Model &Feature &EER \\
		\midrule
		LCNN-BLSTM \cite{zhang21ca_interspeech}&LFCC&16.21 \\
		\midrule
		SELCNN-BLSTM \cite{zhang2021multi}&LFCC&15.93 \\
		\midrule
		LCNN-BLSTM \cite{zhang21ca_interspeech}&W2V2-XLS-R&9.87 \\
		\midrule
		5gMLP \cite{zhang2022partialspoof}&W2V2-Large&9.24 \\
		\midrule
		TDL (w/o ESM)&W2V2-XLS-R&8.79 \\
		\midrule
		TDL &W2V2-XLS-R&\bf 7.04 \\
		\bottomrule
	\end{tabular}
	\label{tab:eerps}
\end{table}
\begin{table}[t]
	\caption{The four evaluation metrics results (\%) for training on 19PS and testing on the LAV-DF test set.}
	\centering
	\setlength{\tabcolsep}{0.5pt}
	\renewcommand{\arraystretch}{0.5}
	\begin{tabular}{cccccc}
		\toprule
		Model &Feature &EER$\downarrow$ &Precision$\uparrow$&Recall$\uparrow$&$F_{1}$ score$\uparrow$ \\
		\midrule
		LCNN-BLSTM&LFCC &17.89&95.93&73.73&83.38 \\
		\midrule
		LCNN-BLSTM &W2V2-XLS-R&15.35&\bf 99.05&62.32&76.50 \\
		\midrule
		TDL &W2V2-XLS-R &\bf 11.23&98.73&\bf 75.42&\bf 85.51 \\
		\bottomrule
	\end{tabular}
	\label{tab:lavdf}
\end{table}
\begin{table}[t]
	\caption{EER results (\%) of different label configuration on 19PS.}
	\centering
	\renewcommand{\arraystretch}{0.5}
	\begin{tabular}{ccccc}
		\toprule
		Label Setting &EER$\downarrow$ &Precision$\uparrow$&Recall$\uparrow$&$F_{1}$ score$\uparrow$  \\
		\midrule
		Boundary 1 &10.89&79.72&82.01&80.85 \\
		\midrule
		real 0 fake 1 &9.52&81.87&84.52&83.17 \\
		\midrule
		real 1 fake 0 &\bf 7.04&\bf 88.69&\bf 95.01&\bf 91.54 \\
		\bottomrule
	\end{tabular}
	\label{tab:label}
\end{table}

\begin{table}[t]
	\caption{Parameters (in thousands) comparison.}
	\centering
	\renewcommand{\arraystretch}{0.2}
	\begin{tabular}{lcc}
		\toprule
		Model& Parameters  \\
		\midrule
		TDL & 8,718 & \\
		\midrule
		LCNN-BLSTM &21,511\\
		\bottomrule
	\end{tabular}
	\label{tab:Parameters}
\end{table}
\subsection{Disscussion}
\noindent \textbf{Label Setting.} As we mentioned in Section 3.2, we set real frames, fake frames for 1 and 0. To the best of our knowledge, there has been no prior research discussing which label configuration will be beneficial to the final prediction. Therefore, we experiment with three different label settings on our proposed TDL model as shown in Table \ref{tab:label}. 

``Boundary 1'' indicates that we set the boundary frames between genuine and fake segments as 1, while other positions are set as 0. In practice, due to the sparsity of boundary frames, we set 4 boundary frames at the transition between genuine and fake segments. Additionally, we employ a weighted BCE loss, assigning a weight value of 100 to the boundary values, as a replacement for standard BCE. Experimental results demonstrate that this method is less effective compared to directly predicting the authenticity of individual frames. Additionally, since predicting boundaries often requires further verification of the genuineness of the segments on both sides, we did not adopt the boundary setting.

For the frame-level direct prediction of authenticity, we conducted experiments by setting real frames as 0 and fake frames as 1, and alternatively by setting real frames as 1 and fake frames as 0, as shown in the “real 0 fake 1” and “real 1 fake 0” of the Table \ref{tab:label} respectively. Experiments results show that “real 1 fake 0” outperform “real 0 fake 1” in four evaluation metrics, especially in recall metric, which indicates that TDL can accurately identify genuine speech. When setting real frames as “1” and fake frames along with padding frames as “0”, we can better concentrate on the real segment. This is similar to previous works \cite{zhang2021one,xie23c_interspeech} which also focus on the real speech distribution in fully-spoofed ADD task. Through our experiments, we have demonstrated that it is also significant in partially-spoofed ADD task. This is also why W2V2 features are effective in the field of ADD which only extracted by rich real source domains. 

\noindent \textbf{Complexity Comparision.} Apart from evaluating the performance, we measured the complexity of the models. For frame-level detection task, particularly for fine-grained prediction, the large final output dimension can result in excessive parameterization and low efficiency. Unlike LCNN, which convolves overall values, our proposed TDL model uses temporal convolution operation to selectively focus only on high-weight regions. It can be observed that the parameter count of TDL is only 40.53\% of that of LCNN-BLSTM, which is shown in Table \ref{tab:Parameters}.

\section{Conclusion}
In this paper, we propose an efficient temporary deepfake location approach based embeddings for partially spoofed audio detection. TDL can achieve outstanding performance benefits from two designed core modules: embedding similarity module and temporal convolution operation, which can effectively capture both feature and positional information. The experimental results demonstrate that TDL achieves the best performance in the 19PS dataset and also perform well in cross-dataset scenarios.

\small
\bibliographystyle{IEEEbib}
\bibliography{strings,refs}
\end{document}